\documentclass{PoS}

\PoS{PoS(LAT2005)287}

\title{Finite volume effects in the gluon propagator}

\ShortTitle{Finite volume effects in the gluon propagator}

\author{\speaker{Orlando Oliveira} and Paulo J. Silva\thanks{Supported by FCT 
via grant SFRH/BD/10740/2002.} \\
Centro de Física Computacional, Departamento de Física, 
Universidade de Coimbra     \\
3004-516 Coimbra, Portugal. \\
E-mail: \email{orlando@teor.fis.uc.pt}, \email{psilva@teor.fis.uc.pt}}

\abstract{We report on a preliminary study of the volume dependence of the 
gluon propagator. The propagator is computed using different lattice volumes, 
its extrapolation to infinite volume is investigated with particular attention
to the its IR behaviour. Our data shows a mild but measurable dependence with
the volume. Unfortunately, we are not able yet to clarify its behaviour close
to zero momentum.}

\FullConference{XXIIIrd International Symposium on Lattice Field Theory\\
                 25-30 July 2005\\
                 Trinity College, Dublin, Ireland}
                
\usepackage{amssymb}
\usepackage{fontenc}
\usepackage{times}
\usepackage{mathptmx}
\usepackage{color}
\usepackage{graphicx}
\usepackage{ifpdf}

\usepackage{psfrag}
\usepackage{bm}

\begin{document}

\section{Introduction and motivation}

The infrared (IR)
properties of the gluon propagator were studied in \cite{OlSi04,SiOl05} 
with the following large assymetric lattices $16^3 \times 128$ and 
$16^3 \times 256$. For these lattices, although the propagator for pure 
temporal momenta are within errors, the equivalent spatial momenta show a 
deviation from the temporal ones for small momenta. Clearly, this is a finite 
volume effect. So the question, to which we will try to give a first and 
preliminary answer, is how reliable are the figures reported in 
\cite{OlSi04,SiOl05}? Or what is the meaning of these results? In order to try
to answer these questions,
we make a first study of the volume dependence of the
gluon propagator, which will also allow an extrapolation to infinite volume.

\section{Lattice setup}

In order to study the volume dependence we generated\footnote{All
configurations were generated with MILC code 
{\tt http://physics.indiana.edu/\~{ }sg/milc.html}.} SU(3) pure gauge, Wilson
action configurations
for the lattices reported in table \ref{Uvol}. The table also 
describes the combined overrelaxed+heat bath Monte Carlo sweeps, the number of
combined sweeps for thermalization, the separation of combined sweeps used
for each simulation and the total number of configurations for each lattice.

\begin{table}[h]
\begin{center}
\begin{tabular}{|r|@{\hspace{0.2cm}}r|
                   @{\hspace{0.2cm}}r|
                   @{\hspace{0.2cm}}r|
                   @{\hspace{0.2cm}}r|}
\hline
   Lattice           & Update
                     & therm.
                     & Sep.
                     & Conf. \\
\hline
   $8^3 \times 256$   & 7OVR+4HB & 1500 & 1000 & 80 \\
   $10^3 \times 256$  & 7OVR+4HB & 1500 & 1000 & 80 \\
   $12^3 \times 256$  & 7OVR+4HB & 1500 & 1000 & 80 \\
   $14^3 \times 256$  & 7OVR+4HB & 3000 & 1000 & 47 \\
   $16^3 \times 256$  & 7OVR+4HB & 3000 & 1500 & 155 \\
\hline
\end{tabular}
\caption{Lattices setup used in the study of the volume dependence.}
\label{Uvol}
\end{center}
\end{table}

For the definition of the gluon propagator, notation and the gauge fixing 
method see \cite{gribovgluon}.

\section{Time \textit{versus} Spatial Momenta}

\begin{figure}[h]
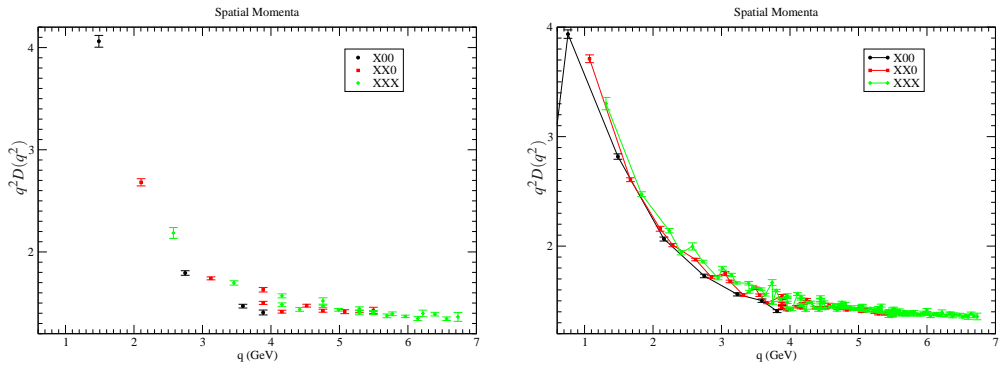

\psfrag{GLUON DRESSING FUNCTION (8.3.256)}{}
\psfrag{GLUON DRESSING FUNCTION (16.3.256)}{}
\psfrag{EIXOX}{ $q(GeV)$}
\psfrag{EIXOY}{\begin{tiny} $q^2 D(q^2)$ \end{tiny}}
   \begin{minipage}[b]{0.45\textwidth}
   \centering
   \includegraphics[origin=c,scale=0.27]{dress_spatial_8.3.256.eps}
   \end{minipage} 
  \begin{minipage}[b]{0.45\textwidth}
  \centering
  \includegraphics[origin=c,scale=0.27]{dress_spatial_16.3.256.eps}
  \end{minipage} 
\caption{Dressing function for different spatial momenta, for the
smaller (left) and largest (right) lattices.}
\label{TSmom}
\end{figure}

Our first check was to look at how the time and the different spatial momenta
evolved with the lattice volume. The gluon propagator, for the smallest
and the largest lattices, for different types of spatial momenta is given in
figure \ref{TSmom}. The data shows that the propagator becomes smaller when
the number of zero components increases. The effect is larger for smaller
momenta. For the largest lattice, the effect is not so strong but is,
nevertheless, clearly visible.

The comparison between temporal and spatial momenta with only one 
nonzero component shows a similar behaviour. The temporal data is
always below the spatial data - see \cite{SiOl05}.

\begin{figure}[h]
\begin{center}
\psfrag{DRESSING FUNCTION T=256}{}
\psfrag{EIXOX}{$q$}
\psfrag{EIXOY}{\begin{tiny}$q^2 D( q^2)$ \end{tiny} }
\vspace{0.1cm}
\includegraphics[origin=c,scale=0.35]{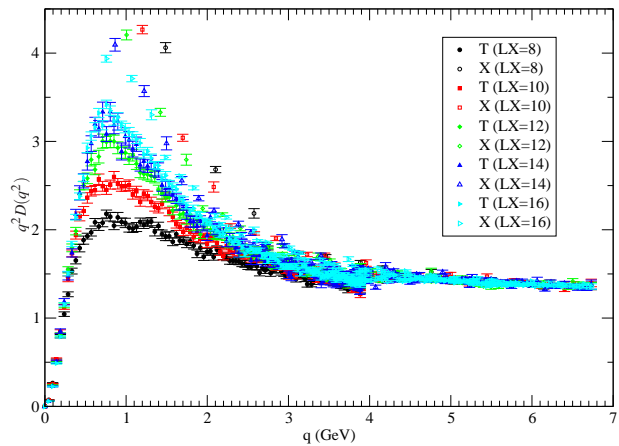}
\caption{Gluon dressing functions}
\label{q2dall}
\end{center}
\end{figure}

The gluon dressing function for temporal momenta and spatial momenta with only
one nonzero component show a strong dependence on the volume,
see figure \ref{q2dall}, specially in the intermediate momenta region
($\sim 1$ GeV). Note that while the temporal data seems to increase with the
volume, the spatial data decreases with the volume. Certainly, this is the
case for almost all the momenta region but not for the lowest momenta - see
figure \ref{DlowV}. Indeed, 
a zoom to the IR region shows the opposite behaviour, i.e. for small momenta
the propagator decreases with the volume. Therefore, if in the IR the gluon
dressing function is given by $( q^2 )^{2 \kappa}$, a decreasing of the 
dressing function means that $\kappa$ should increase with the volume. Then,
the values measured with any finite lattice can be seen, at least, as a lower 
bound in $\kappa$.

\section{IR region}

\begin{figure}[h]
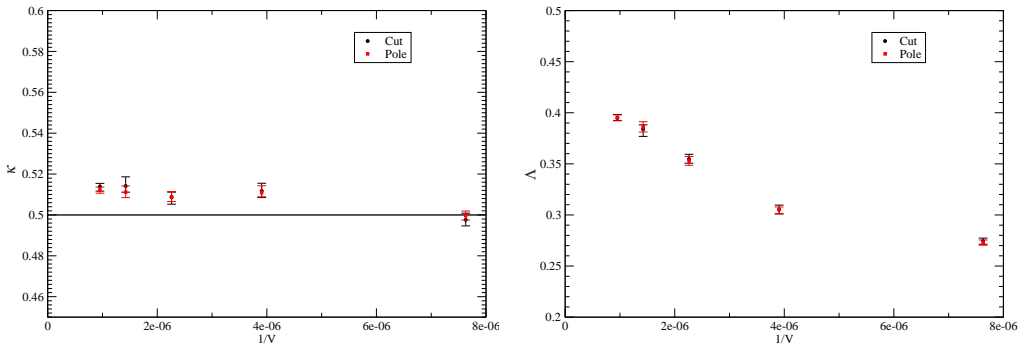

\psfrag{KAPPA IR2 VOLUME}{}
\psfrag{LAMBDA IR2 VOLUME}{}
\psfrag{kappa}{\begin{tiny} $\kappa$ \end{tiny}}
\psfrag{Lambda}{\begin{tiny} $\Lambda$ \end{tiny}}
   \begin{minipage}[b]{0.45\textwidth}
   \centering
   \includegraphics[origin=c,scale=0.27]{k_V_IR2.eps}
   \end{minipage} 
  \begin{minipage}[b]{0.45\textwidth}
  \centering
  \includegraphics[origin=c,scale=0.27]{Lambda_V_IR2.eps}
  \end{minipage} 
\caption{$Z_{cut}$ and $Z_{pole}$ fitted parameters as function of lattice 
volume.}
\label{IRv}
\end{figure}

In \cite{SiOl05} it was seen that the IR region is well described either by
a pole or a cut behaviour\footnote{See \cite{SiOl05} for details.}. We fitted
both expressions to the different lattices. 
As seen in figure \ref{IRv}, there is
good aggreement between the two expressions. Excluding the smallest lattice,
the $\kappa$ (certainly not $\Lambda$) seems to be stable against a change in 
volume. For the largest
volume $\kappa = 0.5138^{+16}_{-22}$ and
$0.5122^{+15}_{-15}$ for the cut and pole
formulas, respectively. In this sense, the lattice data supports a vanishing
zero momentum gluon propagator.

\begin{figure}[h]
\begin{center}
\vspace{0.1cm}
\includegraphics[origin=c,scale=0.4]{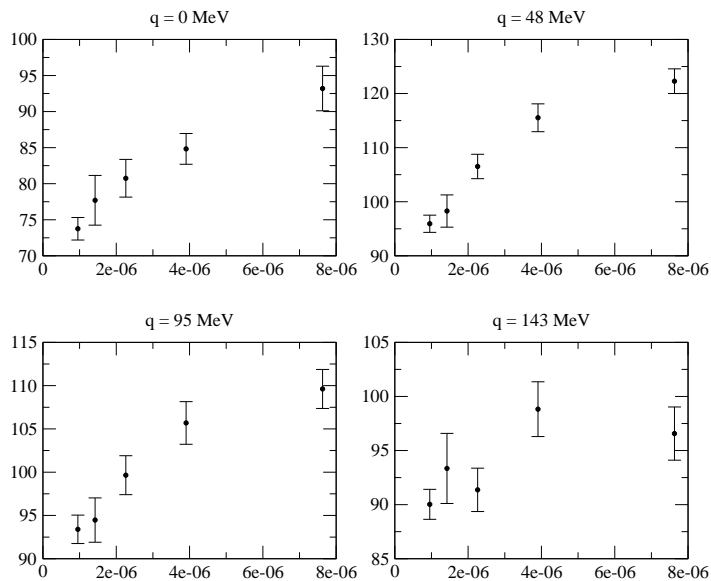}
\caption{Propagator as a function of the inverse volume.}
\label{DlowV}
\end{center}
\end{figure}

Let us now look at the behaviour of the propagator as a function of the 
volume. For the 
four lowest momenta values, the volume dependence can be seen in
figure \ref{DlowV}. To extrapolate the zero momentum gluon propagator we
tried the formulas reported in table 2. The results are a bit
disapointing in the sense that all functional formulas produce good fits.
From the table, one sees that the extrapolation is compatible with all types
of zero momentum behaviour, from zero to infinite. Nevertheless, the results
of table 2 will help to reduce the number of fitting formulas to
be considered in the extrapolation of the full propagator.

\begin{table}[h]
\begin{center}
\begin{tabular}{|lcl|}
\hline
              &  $\chi^2/d.o.f.$  & Fitted Parameters \\
\hline
$a + b x$          & 0.60 & $a = 72.06$, $b = 3.0 \times 10^6$ \\
$a + b x + c x^2$  & 0.24 & $a = 69.52$, $b = 5.1 \times 10^6$,
                                         $c = -2.5 \times 10^{11}$ \\
$a + b x^c$        & 0.10 & $a = 53.96$, $b = 1702$, $c = 0.32$ \\
$a x^b$            & 0.12 & $a = 325.4$, $b = 0.11$ \\
$(a + b x) x^c$    & 0.13 & $a = 2 \times 10^{-5}$, $b = 418$, 
                                  $c = -0.87$ \\
\hline
\end{tabular}
\end{center}
\label{zeroext}
\caption{Zero momenta propagator extrapolation, where $x = 1/V$.}
\end{table}

\section{Infinite volume extrapolation}

\begin{figure}[h]
\begin{center}
\psfrag{EIXOX}{$q$}
\psfrag{EIXOY}{\begin{tiny}$q^2 D( q^2)$ \end{tiny} }
\vspace{0.75cm}
\includegraphics[origin=c,scale=0.35]{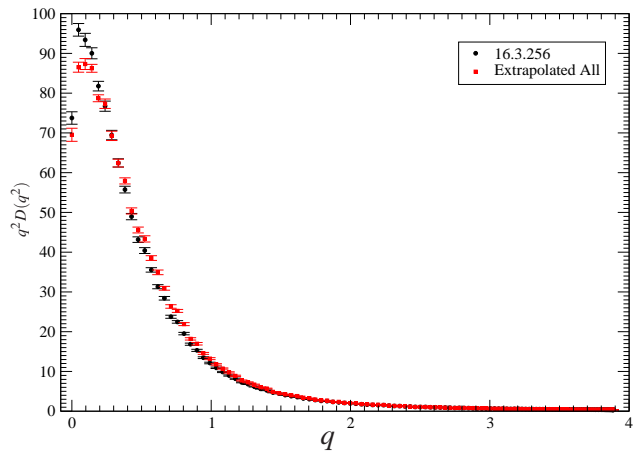}
\caption{Extrapolated gluon propagator.}
\label{Dext}
\end{center}
\end{figure}

As a guide to the infinite volume extrapolation we take the results of the zero
momentum gluon propagator. Since in the last fitting formula 
$a  \sim 10^{-5}$, we will only consider the following fitting functions
\begin{equation}
   a + b x + c x^2, \hspace{1cm} a + b x^c \quad ,
\end{equation}
and will fit only the timelike momenta. Moreover, each timelike momenta will
be fitted separatly. From the two fitting formulas, the second one does not
provide a smooth propagator. For certain momenta the fits do
not converge or produce an infinite propagator. Therefore, from now on we will
consider only the fits assuming a quadratic behaviour. Indeed, for the 
quadratic function, the extrapolated propagator looks rather smooth 
(see figure \ref{Dext}) and only less than 15\% of the fits have a 
$\chi^2/ d. o.f. > 2$.

If one tries to repeat the study performed in \cite{OlSi04,SiOl05} it cames
that the only fits which have acceptable $\chi^2/d.o.f.$ are the fits to
the IR region. Probably, this is due to the inclusion of the data from the
smallest lattice which looks different from all the other ones. The
results of the IR fits are reported in table \ref{Fitse}. Again, the
lattice although pointing towards a vanishing zero momentum,
it does not provide a definite answer. 
Looking at the $\kappa$ values, the fit to the analytical
DSE solution is compatible with 0.5. 
However, it is interesting that if one includes
corrections to the above solution, the result agrees within errors 
with the figures estimated from the largest lattice, using the same procedure.

\begin{table}[h]
\begin{center}
\begin{tabular}{|c|rrr|rrr|}
\hline
          & \multicolumn{3}{c|}{Extrapolated} &
            \multicolumn{3}{c|}{$16^3 \times 256$} \\
          & $\kappa$ & $\Lambda$  & $\chi^2/d.o.f.$      & 
            $\kappa$ & $\Lambda$  & $\chi^2/d.o.f.$      \\
\hline
$\left( q^2 \right)^{2 \kappa}$
          & $0.4993(7)$  &  ---   &  $0.37$   &   
            $0.4858(2)$  &  ---   &  $0.40$ \\
\hline
$\left( q^2 \right)^{2 \kappa} \left( 1 ~ + ~ a \, q^2 \right)$
          & $0.5129(10)$  &  ---   &  $0.07$    &  
            $0.5070(50)$  &  ---   & $0.44$    \\
\hline
$\left( \frac{q^2}{q^2 + \Lambda^2}  \right)^{ 2 \kappa } $
          & $0.5198(2)$   & $438(1)$ & $1.29$  &
            $0.5090(20)$  & $409(4)$ & $0.71$ \\
\hline
$\frac{ ( q^2 )^{ 2 \kappa } }{ ( q^2 )^{ 2 \kappa } +  ( q^2 )^{ 2 \kappa }}$
          & $0.5167(1)$   & $439(1)$ & $1.36$ &
            $0.5077(17)$  & $409(4)$ & $0.69$ \\
\hline
\end{tabular}
\caption{IR fits. The errors on the extrapolated propagator are clearly
underestimated.}
\label{Fitse}
\end{center}
\end{table}

\section{Conclusions}

We have performed a first study of the volume dependence of the gluon
propagator. Although the lattice data seems to favour $\kappa \geq 0.5$,
we are not able yet to provide a clear answer concerning the behaviour
of the gluon propagator at zero momentum. 
In what concerns the IR region, our data shows a mild but measurable
dependence with the volume. This is not in aggreement with a recent similar 
study of the DSE equations on a tours \cite{Fi05,Fi005}.

\section*{Acknowledgements}

We would like to thank A. C. Aguillar, C. S. Fischer, A. G. Williams,
J. I. Skullerud, P. Bowman and D. Leinweber for fruitful and inspiring 
discussions.

\end{document}